\documentclass[pra,twocolumn,aps,showpacs,superscriptaddress]{revtex4-1}
\usepackage{bm}%
\usepackage{float}
\usepackage[colorlinks=true,linkcolor=blue]{hyperref}%
\expandafter\ifx\csname package@font\endcsname\relax\else
 \expandafter\expandafter
 \expandafter\usepackage
 \expandafter\expandafter
 \expandafter{\csname package@font\endcsname}%
\fi

\usepackage{array}
\usepackage{amsmath,amssymb}
\usepackage{MnSymbol}
\usepackage{tikz-feynman} 
\usepackage{graphicx}
\usepackage{mathrsfs}
\usepackage{bm}
\usepackage{mathtools}
\usepackage{epstopdf}
\usepackage{hyperref}

\hypersetup{%
   pdfpagemode=None, %FullScreen,
   pdfstartpage=1,
   pdfmenubar=true,
   pdftoolbar=true,
   colorlinks = true,
   linkcolor=blue,
   citecolor=blue,
   urlcolor=blue,
   bookmarksopen=false
 }
\usepackage{amsmath}
\usepackage{amsfonts}
\usepackage{mathtools}
\usepackage{graphicx}
\usepackage{fixme}
\usepackage{color}
\usepackage{soul}
\usepackage{ulem}
\usepackage{dcolumn}
\usepackage{bm}

\usepackage{tikz}
\usepackage{tikz-feynman}
\tikzfeynmanset{compat=1.1.0}

\begin{document}

\title{Exceptional points in dissipative coupling polaron-polaritons}

\author{A. J. Vega-Carmona}
\affiliation{%
Departamento de Física, Universidad Autónoma Metropolitana-Iztapalapa,
Av. Ferrocarril San Rafael Atlixco 186, C.P. 09310 CDMX, Mexico}%
\author{D. A. Mendoza}
\affiliation{%
Departamento de Física, Universidad Autónoma Metropolitana-Iztapalapa,
Av. Ferrocarril San Rafael Atlixco 186, C.P. 09310 CDMX, Mexico}%
\author{A. Camacho-Guardian}
\affiliation{%
Instituto de Física, Universidad Nacional Autónoma de México, Apartado Postal 20-364, Ciudad de México, Código Postal 01000, Mexico}%
\author{M. A. Bastarrachea-Magnani}
\affiliation{%
Departamento de Física, Universidad Autónoma Metropolitana-Iztapalapa,
Av. Ferrocarril San Rafael Atlixco 186, C.P. 09310 CDMX, Mexico}%

%%%%%%%%%%%%%%%%%%%%%%%%%%%%%%%%%%%%%%%%%%%%%%%%%%%%%%%%%%%%
%%%%%%%%%%%%%%%%%%%%%%%%%%%%%%%%%%%%%%%%%%%%%%%%%%%%%%%%%%%%
\begin{abstract}
Understanding how strong correlations and dissipation combine to shape collective quantum excitations is a central challenge in many-body physics. We investigate the effect of dissipative light-matter coupling on strongly interacting exciton-polaritons in the presence of a biexciton resonance, which gives rise to polaron-polariton quasiparticles. We show that the interplay between many-body correlations and non-Hermitian coupling generates anomalous dispersion relations and exceptional points in the polaron-polariton spectrum. The location and coexistence of exceptional points are controlled by the dissipative coupling and the relative decay rates of the excitonic and photonic constituents, allowing them to emerge across different polaron-polariton branches. These results identify dissipative polaron-polaritons as a versatile platform for exploring non-Hermitian many-body physics with tunable light-matter quasiparticles.
\end{abstract}
%%%%%%%%%%%%%%%%%%%%%%%%%%%%%%%%%%%%%%%%%%%%%%%%%%%%%%%%%%%%
%%%%%%%%%%%%%%%%%%%%%%%%%%%%%%%%%%%%%%%%%%%%%%%%%%%%%%%%%%%%

\maketitle

%%%%%%%%%%%%%%%%%%%%%%%%%%%%%%%%%%%%%%%%%%%%%%%%%%%%%%%%%%%%
%%%%%%%%%%%%%%%%%%%%%%%%%%%%%%%%%%%%%%%%%%%%%%%%%%%%%%%%%%%%
\section{Introduction}
%%%%%%%%%%%%%%%%%%%%%%%%%%%%%%%%%%%%%%%%%%%%%%%%%%%%%%%%%%%%
%%%%%%%%%%%%%%%%%%%%%%%%%%%%%%%%%%%%%%%%%%%%%%%%%%%%%%%%%%%%

A defining challenge in quantum physics is to understand how concepts rooted in linear, single-particle descriptions evolve when interactions and losses are treated on equal footing. While many quantum many-body systems---ranging from mode hybridization to collective excitations---can be understood within Landau’s quasiparticle framework~\cite{landau1957theory,abrikosov1959theory}, even in open settings, recent experiments have demonstrated that dissipation can qualitatively reorganize the excitation spectrum itself, giving rise to collective modes governed by non-Hermitian rather than Hermitian physics~\cite{ashida2020non}. In systems with dissipative coupling~\cite{Harder2021,Wang2020}, the emergence of level attraction, anomalous dispersions~\cite{Wurdack2023,Bleu2024}, and exceptional points (EPs)~\cite{Heiss_2012,Miri2019} signals a departure from equilibrium paradigms, reshaping quasiparticles in ways with no analog in closed quantum systems.

Exciton-polaritons in semiconductors provide a versatile platform to explore such emergent quasiparticles arising from the interplay of interactions and losses. Strong polariton-polariton interactions have been theoretically predicted and experimentally realized through polaritonic Feshbach resonances~\cite{Wouters2007,Takemura2014,Takemura2017,Carusotto_2010,Kumar2023}, Rydberg nonlinearities~\cite{gu2021enhanced,makhonin2024nonlinear,walther2018giant}, and moir\'e-induced interactions~\cite{Zhang2021,Park2023,Camacho2022,du2024nonlinear,HerreraGonzalez2025}. In particular, polaritonic Feshbach resonances have enabled the realization of strongly interacting polaron-polaritons, mediated by an underlying bound state that can be either a biexciton~\cite{Carusotto_2010,Borri2000,Saba2000,Baars2001} or a trion~\cite{Efimkin2017,Li2021}. Depending on whether the host medium is bosonic~\cite{Takemura2014,Levinsen2019} or fermionic~\cite{sidler2017fermi,Kumar2023}, the resulting spectrum realizes Bose or Fermi polaron-polaritons. These systems have enabled electrical control and acceleration of polaritons~\cite{Cotle2019,Chervy2020,Myers2025}, strong mediated interactions~\cite{emmanuele2020highly,Tan2020,Tan2023,Bastarrachea2021,Camacho2021,muir2022interactions}, and proposals for quantum many-body phases ranging from conventional superfluidity~\cite{Laussy2010,Cotle2016,Choo2025} to topological superconductivity~\cite{Julku2022}, supersolidity~\cite{Shelykh2010}, quantum droplets~\cite{Caldara2026}, among others.

Despite these advances, strongly interacting polaritons have predominantly been explored within frameworks that remain effectively Hermitian. Only a few studies have addressed how non-Hermitian coupling can fundamentally reorganize polaritonic quasiparticles, leading to phenomena such as Zeno effects~\cite{wasak2021quantum} or dissipation-induced anomalous dispersion near trion resonances~\cite{dhara2018anomalous}. In particular, EPs are spectral singularities in which eigenvalues coalesce, granting them special topological properties~\cite{Heiss2001,Miri2019,Downing2025}, and dramatic effects on transport and scattering~\cite{Heiss2010}. They have attracted attention in recent years for their ubiquitous nature~\cite{Heiss2012} and potential applications for quantum technologies such as gain-loss manipulation~\cite{Li2022,Wingenbach2024} or high-sensitivity sensors~\cite{Chen2017}. The high tunability of exciton-polariton systems makes them a promising platform to exploit applications of EPs~\cite{Gao2015,Gao2018,Hanai2019,Yu2021,Liao2021,Kwong2026arXiv}. 

In this work, we show that dissipative light--matter coupling qualitatively transforms the polaron-polariton spectrum, generating EPs involving distinct quasiparticle branches and inducing anomalous dispersion relations. Our results identify dissipative polaron-polaritons as a new class of collective excitations, in which many-body correlations and non-Hermitian physics are intrinsically intertwined.

In Sec.~\ref{sec:2} we present the model and discuss the many-body theory. Next, in Sec.~\ref{sec:3}, we recall the polariton states arising from the inclusion of dissipative coupling. In Sec.~\ref{sec:4} we extend the study to the formation of polaron-polaritons under dissipative coupling, and analyze the emergence of exceptional points over the parameter space. Finally, in Sec.~\ref{sec:5} we discuss our results and present our conclusions. We include an Appendix with complementary calculations of the many-body theory.

%%%%%%%%%%%%%%%%%%%%%%%%%%%%%%%%%%%%%%%%%%%%%%%%%%%%%%%%%%%%
%%%%%%%%%%%%%%%%%%%%%%%%%%%%%%%%%%%%%%%%%%%%%%%%%%%%%%%%%%%%
\section{Hamiltonian and many-body theory}
\label{sec:2}
%%%%%%%%%%%%%%%%%%%%%%%%%%%%%%%%%%%%%%%%%%%%%%%%%%%%%%%%%%%%
%%%%%%%%%%%%%%%%%%%%%%%%%%%%%%%%%%%%%%%%%%%%%%%%%%%%%%%%%%%%

We consider a mixture of two species of exciton-polariton generated by circularly polarized light $\sigma=\{\uparrow,\downarrow\}$ in a microcavity semiconductor. The $\downarrow$ exciton-polaritons form a BEC and constitute the majority species, while there is a low concentration of the $\uparrow$ exciton-polaritons such that they are regarded as impurities. The system's Hamiltonian reads
\begin{gather}    \hat{H}=\hat{H}_{0}+\hat{H}_\mathrm{lm}+\hat{H}_\mathrm{int},
\end{gather}
where
\begin{gather}
\hat{H}_{0}=\sum_{\sigma,\mathbf{k}}\left[\left(\epsilon^{x}_{\mathbf{k}}-i\gamma^{x}\right)\hat{x}^\dagger_{\sigma,\mathbf{k}}\,\hat{x}_{\sigma,\mathbf{k}}+\right.
%\\ \nonumber
\left. \left(\epsilon^{c}_{\mathbf{k}}-i\gamma^{c}\right)\hat{c}^\dagger_{\sigma,\mathbf{k}}\,\hat{c}_{\sigma,\mathbf{k}}\right]
\end{gather}
is the non-interacting term. Here $\hat{x}^\dagger_{\sigma,\mathbf{k}}$ and $\hat{c}^\dagger_{\sigma,\mathbf{k}}$ are the creation operators of excitons and cavity photons, respectively, with polarization $\sigma$ and momentum $\mathbf{k}$. Their energy dispersion are $\epsilon^x_{\mathbf{k}}=\mathbf{k}^2/2m_x$ and $\epsilon^c_{\mathbf{k}}=\delta+\mathbf{k}^2/2m_c$, respectively, where $m_x$ and $m_c$ are the effective exciton and cavity photon masses, and $\delta$ is the light-matter detuning at zero momentum. We assume both the cavity photons and excitons are interacting with their own environment, so we include their own decay rates given by $\gamma^{x}$ and $\gamma^{c}$. For photons this comes from losses due to the imperfection of the cavity mirrors; for excitons decay comes from interactions with lattice vibrations, imperfections and thermal effects. The light-matter interaction is
\begin{gather}
\hat{H}_\mathrm{lm}=
   \left(\Omega_\mathrm{Re}-i\Omega_{\mathrm{Im}}\right)\sum_{\sigma,\mathbf{k}}\left(\hat{x}^\dagger_{\sigma,\mathbf{k}}\hat{c}_{\sigma,\mathbf{k}}+\hat{c}^\dagger_{\sigma,\mathbf{k}}\hat{x}_{\sigma,\mathbf{k}}\right),
\end{gather}
where the light-matter coupling is described by a complex term, with $\Omega_\mathrm{Re}$ being the coherent coupling and $\Omega_\mathrm{Im}$ the dissipative coupling, both of which we assume are independent of the momentum and polarization. Here, we regard dissipative coupling as independent of the decay rates $\gamma^{x,c}$ by considering that there is an external common environment for light and matter. We consider the problem under the rotating-wave approximation (RWA) and the assumption of the strong-coupling light-matter regime in the sense that the coherent interaction is larger than the sum of the loss rates~\cite{Novotny2010,Rodriguez2016}. However, we are considering situations in which the individual photon and exciton decay rates might be large to complete the parametric variation picture.

Finally, we include an inter-species polariton interaction via a two-body term
\begin{gather}
    \hat{H}_\mathrm{int}=\frac{g}{2}\sum_\mathbf{k} \hat{x}^\dagger_{\uparrow,\mathbf{k}+\mathbf{q}}\,\hat{x}^\dagger_{\downarrow,\mathbf{k}'-\mathbf{q}}\,\hat{x}_{\downarrow,\mathbf{k}'}\,\hat{x}_{\uparrow,\mathbf{k}}
\end{gather}
where we consider the exciton-exciton interaction to be momentum-independent, with $g$ denoting the two-body interaction strength, which will be expressed in terms of the binding energy $\epsilon_{B}$ of a biexciton state formed by the $\uparrow$ and $\downarrow$ exciton pair~\cite{Wouters2007,Carusotto2010}.

%%%%%%%%%%%% 
\begin{center}
\begin{figure}[h!]
\includegraphics[width=0.5 \textwidth]{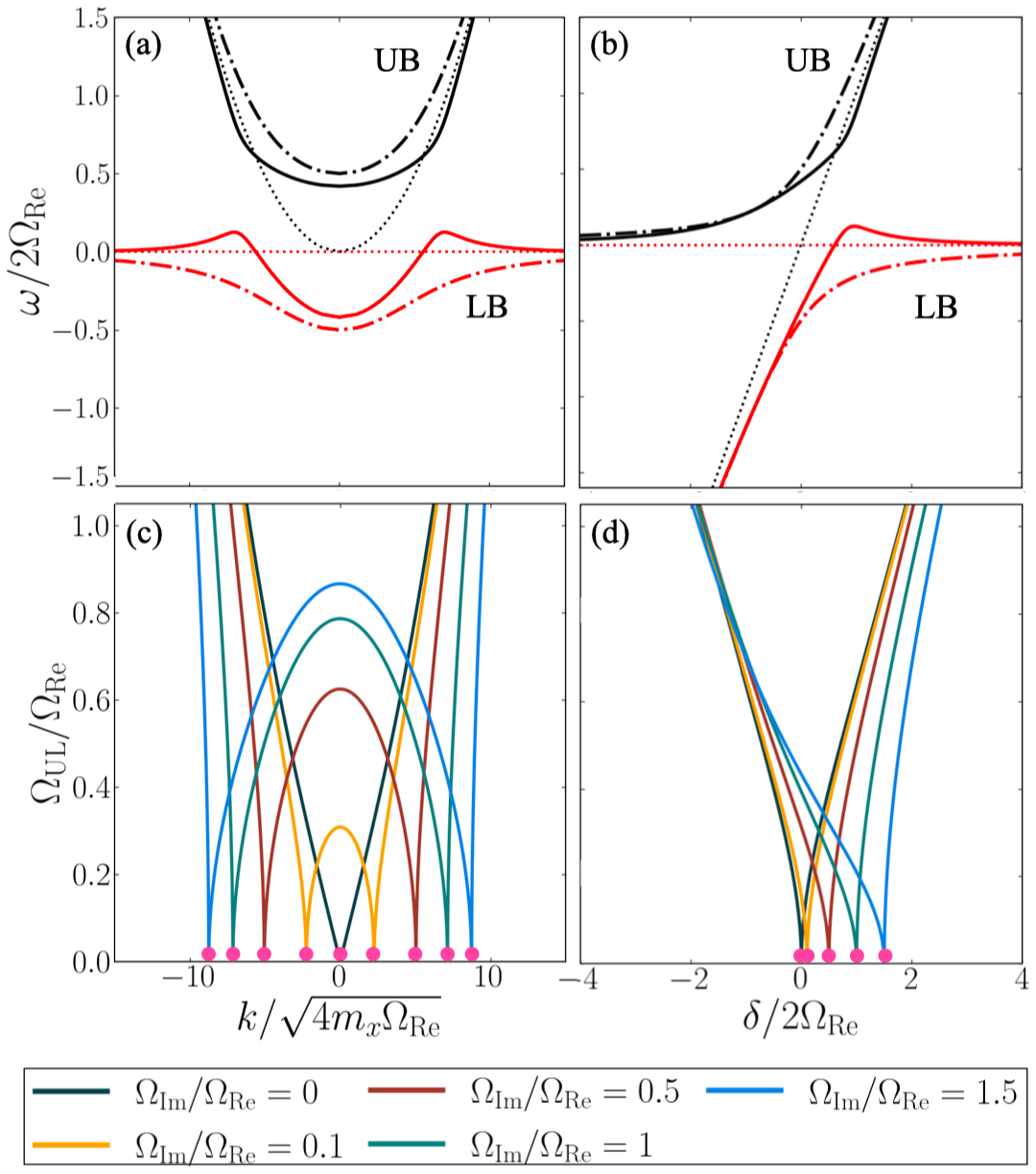} 
\caption{Dissipative coupling polariton branches. As a function of (a) momentum at zero detuning and (b) zero-momentum light-matter detuning. The upper branch (UB) and lower branch (UB) are indicated by black and red curves, respectively. Here, $\Omega_\mathrm{Im}/\Omega_{Re}=0.9$. Below, the condition for the emergence of EPs is shown as a function of dissipative coupling and (c) momentum and (d) detuning for several values of $\Omega_\mathrm{Im}/\Omega_{\mathrm{Re}}$ whose colors are indicated in the figure. The EPs are indicated with pink dots. We employ $\gamma_c/2\Omega_\mathrm{Re}=0.1$ and $\gamma_x/2\Omega_\mathrm{Re}=0.9$.}
\label{fig:1}
\end{figure}
\end{center}
%%%%%%%%%%%% 

%%%%%%%%%%%%%%%%%%%%%%%%%%%%%%%%%%%%%%%%%%%%%%%%%%%%%%%%%%%%
\subsection{Many-body theory}
%%%%%%%%%%%%%%%%%%%%%%%%%%%%%%%%%%%%%%%%%%%%%%%%%%%%%%%%%%%%

To describe the properties of the system, we employ the finite-temperature Green's function formalism. The exciton-polariton system in the absence of strong exciton-exciton interactions can be represented through a $2\times2$ propagator matrix, which in Matsubara frequency space reads~\cite{Bastarrachea2019},
\begin{gather}
\mathbf{G}^{-1}(k)=\mathbf{G}_{0}^{-1}(k)-\mathbf{\Sigma}(k)
\end{gather}
where $k=(\mathbf{k},i\omega_n)$ is the energy-momentum quadrivector, $i\omega_n$ are bosonic Matsubara frequencies, 
\begin{gather}
    \mathbf{G}_{0}^{-1}(k)=\begin{bmatrix}
       G_{0\uparrow}^{-1}(k) & 0 \\
        0 & G_{0c}^{-1}(k)
    \end{bmatrix},
\end{gather}
is the free propagator, with $G_{0\uparrow}^{-1}(k)= i\omega_n-\xi^{x}_{\mathbf{k}}+i\gamma^x$, $G_{0c}^{-1}(k)= i\omega_n-\xi^{c}_\mathbf{k}+i\gamma^{c}$ are the free minority exciton and cavity photon propagators. with $\xi_{\sigma,\mathbf{k}}^{x}=\epsilon_{\mathbf{k}}^{x}-\mu_{\sigma}$ where $\mu_{\sigma}$ is the chemical potential of the $\sigma$ exciton. Interactions are included via a $2\times2$ self-energy
\begin{gather}
    \mathbf{\Sigma}(k)=\begin{bmatrix}
       0 & \Omega_\mathrm{Re}-i\Omega_\mathrm{Im} \\
        \Omega_\mathrm{Re}-i\Omega_\mathrm{Im} & \Sigma_{\uparrow}(k)
    \end{bmatrix}.
\end{gather}
where the many-body effects are enconded in the exciton self-energy $\Sigma_{\uparrow}(k)=n_{\downarrow}\mathcal{T}(k)$, being $n_{\downarrow}$ is the density of the BEC, $\mathcal{T}^{-1}(k)=g^{-1}-\Pi_{\uparrow\downarrow}(k)$ is the T-matrix, and $\Pi_{\uparrow\downarrow}(k)$ the pair propagator as described in App.~\ref{app:1}. The polariton spectrum is obtained by performing the analytic continuation $\mathbf{G}(\mathbf{k},\omega)=\mathbf{G}(\mathbf{k},i\omega_n)|_{i\omega_n\to \omega+i0^+}$. Then, one looks for the poles via the determinant of the propagator $\det[\mathbf{G}^{-1}(\mathbf{k},\omega)]=0$. The propagator has an analytical expression in terms of the self-energy, which reads
\begin{gather}
\Sigma_{\uparrow}(\mathbf{k},\omega)=\frac{4\pi n_{\downarrow}}{m_{x}}
\frac{1}{\ln\left(\frac{|\epsilon_{B}|}{-\omega+\frac{\mathbf{k}^{2}}{4m_{x}}}\right)-i\pi\Theta\left(\omega-\frac{\mathbf{k}^{2}}{4m_{x}}\right)}   
\end{gather}

%%%%%%%%%%%%
%%%%%%%%%%%%%%%%%%%%%%%%%%%%%%%%%%%%%%%%%%%%%%%%%%%%%%%%%%%%
%%%%%%%%%%%%%%%%%%%%%%%%%%%%%%%%%%%%%%%%%%%%%%%%%%%%%%%%%%%%
\section{Dissipative coupling polaritons}
\label{sec:3}
%%%%%%%%%%%%%%%%%%%%%%%%%%%%%%%%%%%%%%%%%%%%%%%%%%%%%%%%%%%%
%%%%%%%%%%%%%%%%%%%%%%%%%%%%%%%%%%%%%%%%%%%%%%%%%%%%%%%%%%%%

Before including interactions between polaritons of different polarizations, we first recall the fundamental system of exciton polaritons in the strong-coupling regime. 

%%%%%%%%%%%%%%%%%%%%%%%%%%%%%%%%%%%%%%%%%%%%%%%%%%%%%%%%%%%%
\subsection{Non-interacting case}
%%%%%%%%%%%%%%%%%%%%%%%%%%%%%%%%%%%%%%%%%%%%%%%%%%%%%%%%%%%%

In the ideal case, i.e., in the absence of dissipation sources ($\gamma^x=\gamma^c=\Omega_{\text{Im}}=0$) and with vanishing polariton interactions, one gets the standard lower and upper polaritons~\cite{Hopfield1958} $\epsilon^{\mathrm{UP}/\mathrm{LP}}_\mathbf{k}=(1/2)\left(\delta_\mathbf{k}+2\epsilon^x_\mathbf{k}\pm\sqrt{\delta^2_\mathbf{k}+4\Omega_\mathrm{Re}^2}\right)$, where $\delta_\mathbf{k}=\epsilon^c_\mathbf{k}-\epsilon^x_\mathbf{k}$ is the momentum-dependent detuning. Including all sources of dissipation leads to complex polariton branches of the form~\cite{Mendoza2025arXiv}
\begin{gather}
\nonumber    E_{\mathbf{k}}^{\text{UP}/\text{LP}}=\epsilon_\mathbf{k}^{\mathrm{UP}/\mathrm{LP}}-i\gamma_\mathbf{k}^{\mathrm{UP}/\mathrm{LP}}=\dfrac{1}{2}\left[\delta_\mathbf{k}+ 2\epsilon^x_\mathbf{k}-i\left(\gamma^c+\gamma^x\right) \pm \right. \\
    \pm \left. \sqrt{\left(\delta_\mathbf{k}-i\Delta\gamma\right)^2+4(\Omega_\mathrm{Re}-i\Omega_\mathrm{Im})^2}\right].
\end{gather}
where $\Delta\gamma=\gamma^c-\gamma^x$ is the relative decay. Figure~\ref{fig:1} shows the real part of the upper and lower polariton branches as a function of momentum [panel (a)] and detuning [panel (b)], without and with sources of dissipation. For the latter case, we set the cavity photon decay to $\gamma_c/2\Omega_\mathrm{Re}=0.1$, the exciton decay to $\gamma_x/2\Omega_\mathrm{Re}=0.9$, and the dissipative coupling to $\Omega_\mathrm{Im}/\Omega_\mathrm{Re}=0.9$, where the values have been chosen to make the effect more noticeable. In Fig.~\ref{fig:1} (a), we observe that dissipative coupling gives rise to a curvature change of the dispersion relation, leading to negative mass effects, as it has been experimentally observed before~\cite{Wurdack2023}. Likewise, in Fig.~\ref{fig:1} (b) we notice that the presence of the decay rates is responsible for the existence of level attraction~\cite{Harder2018,Persson2000}, i.e., the reduction of the energy gap between the upper and lower polariton branches, accompanied by a shift toward positive detuning values. Level attraction ultimately leads to exceptional points, i.e., the coalescence of the branches. The EPs can be found from the condition $E_{\mathbf{k}}^{\text{LP}}=E_{\mathbf{k}}^{\text{UP}}$, or~\cite{Mendoza2025arXiv}
\begin{gather} 
\delta_{\textbf{k}}^{\text{EP}}=\pm2\Omega_{\text{Im}},\,\,\,\,\Delta\gamma=\mp2\Omega_{\text{Re}}.
\end{gather}
As a result, in momentum space, two EPs emerge on both sides of the dispersion curves shown in Figs.~\ref{fig:1} (c). Also, they are shifted from the maximum light-matter coupling as a function of detuning, as shown in Figs.~\ref{fig:1} (d), for different values of the dissipative coupling.  

%%%%%%%%%%%% 
\begin{figure}[h!]
\centering
\includegraphics[width=0.49\textwidth]{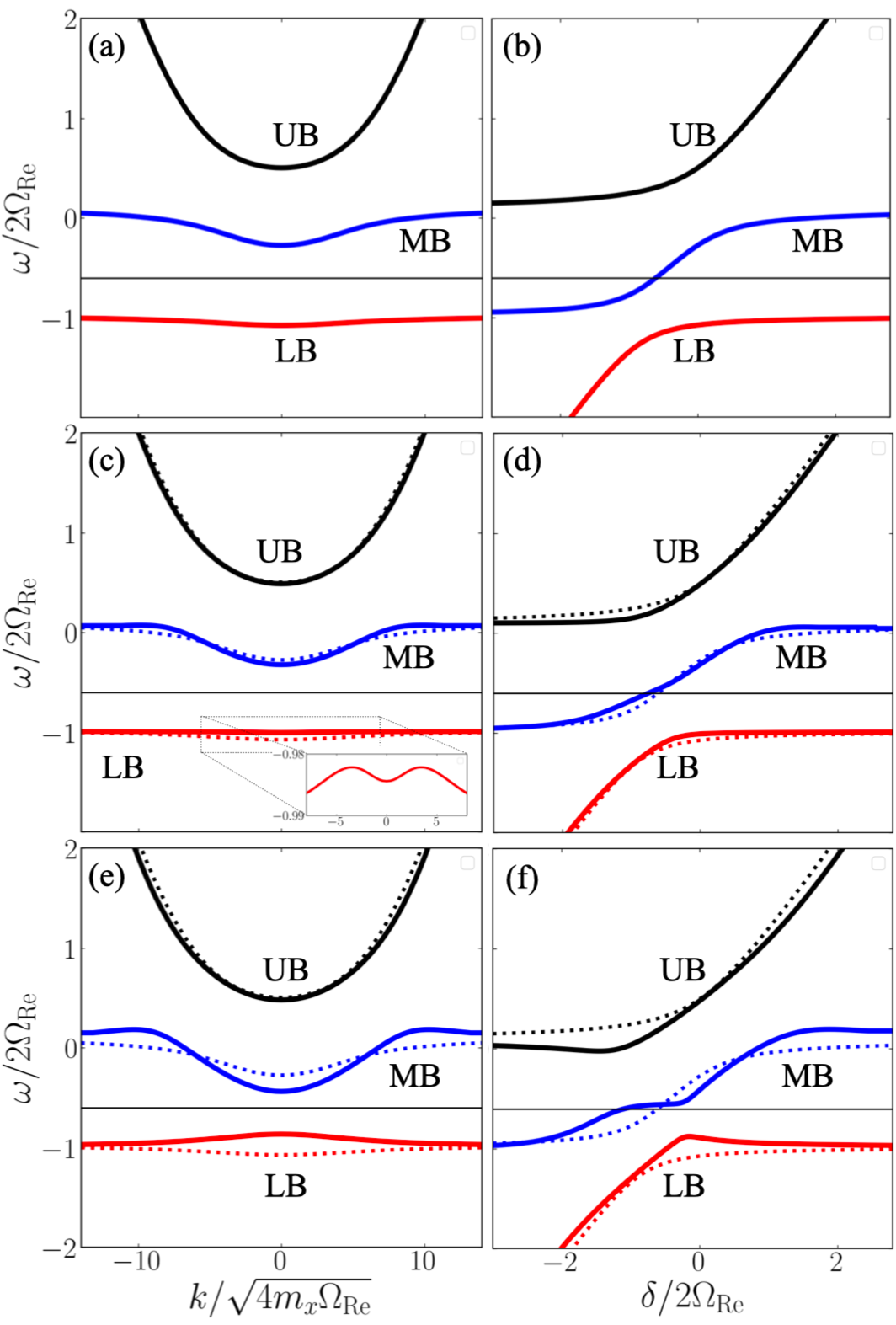}
\caption{Polaron-polaritons as a function of momentum (left) and detuning (right) for increasing values of the dissipative coupling: (a)-(b) $\Omega_{\mathrm{Im}}/\Omega_\mathrm{Re}=0.1$, (c)-(d) $\Omega_{\mathrm{Im}}/\Omega_\mathrm{Re}=1.0$, and (e)-(f) $\Omega_{\mathrm{Im}}/\Omega_\mathrm{Re}=1.5$. In all cases we consider $\gamma^{c}/2\Omega_{\text{Re}}=0.1$, $\gamma^{x}/2\Omega_{\text{Re}}=0.2$, and a BEC density of $4\pi n_{\downarrow}/m_{x}=0.5$. The dissipative coupling UB, MB, and LB are indicated by black, blue, and red solid lines, respectively. The dotted curves correspond to the results in (a) and (b). A black dashed line depicts the biexciton binding energy.}
\label{fig:2}
\end{figure}
%%%%%%%%%%%%

%%%%%%%%%%%%%%%%%%%%%%%%%%%%%%%%%%%%%%%%%%%%%%%%%%%%%%%%%%%%
%%%%%%%%%%%%%%%%%%%%%%%%%%%%%%%%%%%%%%%%%%%%%%%%%%%%%%%%%%%%
\section{Dissipative polaron-polaritons}
\label{sec:4}
%%%%%%%%%%%%%%%%%%%%%%%%%%%%%%%%%%%%%%%%%%%%%%%%%%%%%%%%%%%%
%%%%%%%%%%%%%%%%%%%%%%%%%%%%%%%%%%%%%%%%%%%%%%%%%%%%%%%%%%%%

In this section, we include the exciton-exciton strong interactions via a biexciton resonance. Feshbach physics results in a new light-matter quasiparticle: the polaron-polaritons. Next, we analyze how dissipative coupling parametrically impacts the polaron-polariton branches.

\subsection{Polaron-polariton branches}

In the presence of exciton-exciton interactions, we obtain the polariton branches by looking for the zeros of the real part of the propagator's determinant $\det[\mathbf{G}(\mathbf{k},\omega)]=0$. As a result, the polariton branches emerge from the solution of the transcendental equation 
\begin{gather}\label{eq: pp dr}
    E_{\mathbf{k}}^{\mu}=\frac{1}{2}\bigg(\epsilon^x_\mathbf{k}+\epsilon^c_\mathbf{k}+\Sigma_{\uparrow}(\mathbf{k},E_\mathbf{k}^{\mu})-i\left(\gamma^x+\gamma^c\right)\pm\\ \nonumber
    \pm\sqrt{\left(\delta_{\mathbf{k}}-i\Delta\gamma-\Sigma_{\uparrow}(\mathbf{k},E_\mathbf{k}^{\mu})\right)^2+4(\Omega_\mathrm{Re}-i\Omega_\mathrm{Im})^2}\,\bigg).
\end{gather}
where $\mu$ stands for the polariton branch. In the absence of dissipative coupling, Eq.~\ref{eq: pp dr} leads to three polariton branches arising from the many-body effects of the impurity moving through the interacting medium. These branches stem from the strong material interactions and are known as the lower (LB), middle (MB), and upper (UB) polaron-polariton branches~\cite{Bastarrachea2019}.

%%%%%%%%%%%%
\begin{figure}[h!]
\centering
\includegraphics[width=0.49
\textwidth]{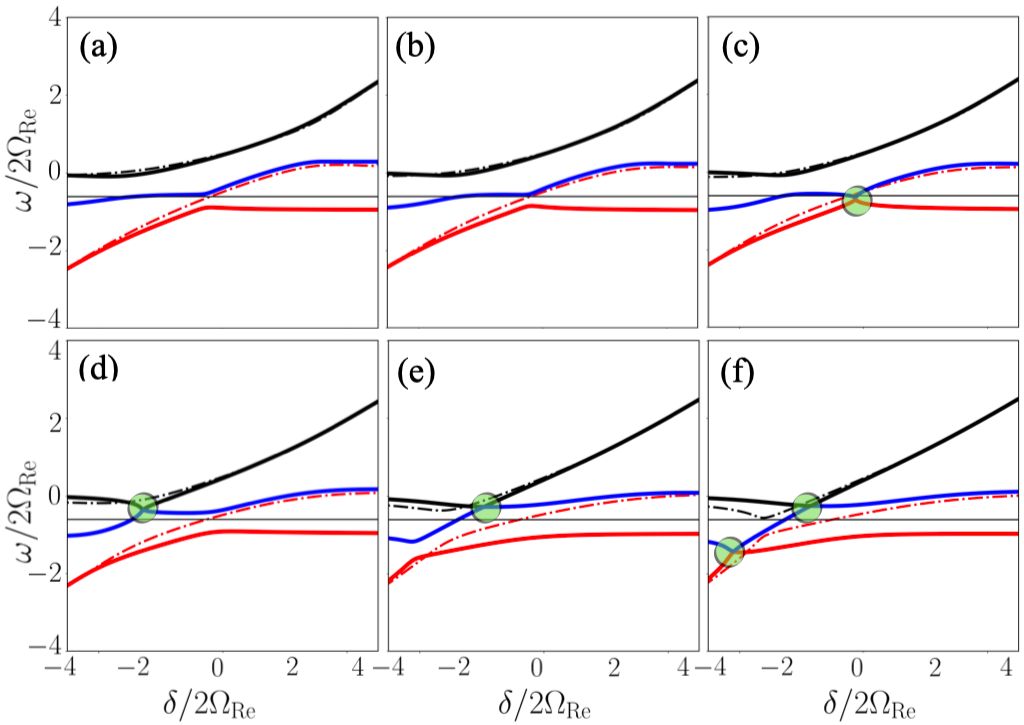}
\caption{Onset of EPs in polaron-polaritons as a function of detuning for zero momentum. The values taken for $\Delta \gamma$ are (a) $-0.4$, (b) $-0.2$, (c) $0$, (d) $0.3$, (e) $0.7$, and (f) $0.9$. We set $\Omega_\mathrm{Im}/\Omega_\mathrm{Re}=1.6$. The color coding is the same as in Fig.~\ref{fig:2}. EPs are emphasized with green circles.}
\label{fig:3}
\end{figure}
%%%%%%%%%%%%

%%%%%%%%%%%%
\begin{figure*}
\centering
\includegraphics[width=1\textwidth]{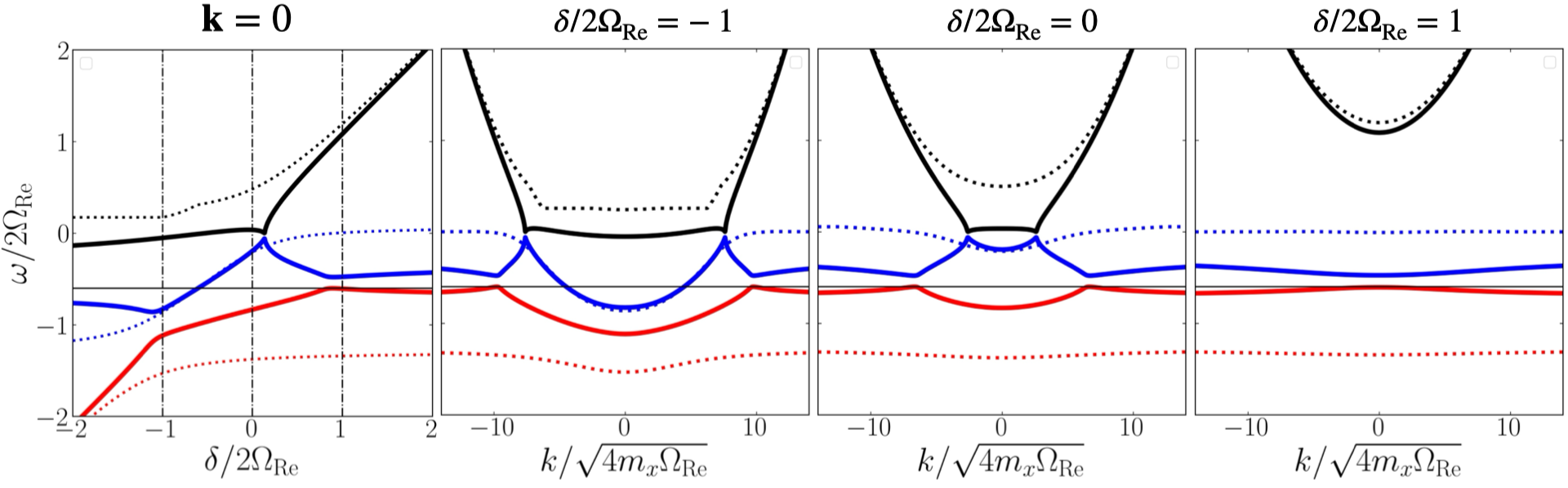}
\caption{Polaron polaritons as a function of detuning for a BEC density of $4\pi n_\downarrow /m_x=1.0$. We employ $\gamma_c/2\Omega_\mathrm{Re} = 0.1$, $\gamma_x/2\Omega_\mathrm{Re}=1.7$ and $\Omega_\mathrm{Im}/\Omega_\mathrm{Re}=0.1$. The color coding is the same as in Fig.~\ref{fig:2}.}
\label{fig:4}
\end{figure*}
%%%%%%%%%%%%

We show the polaron-polaritons’ dispersion relation of the three polaron-polariton branches, where we identify the LB (red solid line), MB (blue solid line), and UB (black solid line) in Fig.~\ref{fig:2}. We employ a BEC density of $n_{\downarrow}=4.6\times10^{10} \mathrm{cm}^{-2}$ ($4\pi n_{\downarrow}/m_{x}=0.5$) and $\epsilon_{B}/2\Omega=-0.6$, as a working example. For comparison, in panels (a) and (b), we present the standard case without non-Hermitian features as a function of momentum and detuning, respectively. The branches are associated with two avoided crossings that result from the coupling between light and emerging matter quasiparticles: the attractive polaron and the repulsive polaron~\cite{Massignan2014}.

Next, in Figs.~\ref{fig:2} (c)-(f), we include the exciton and photon decays and dissipative light-matter coupling setting them to $\gamma^{c}/2\Omega_{\text{Re}}=0.1$, $\gamma^{x}/2\Omega_{\text{Re}}=0.2$ and $\Omega_\mathrm{Im}/\Omega_\mathrm{Re}=1.0$, respectively. We observe changes in the dispersion relation, as it has been experimentally demonstrated before~\cite{dhara2018anomalous,Wurdack2023,Bieganska2024}. In Fig.~\ref{fig:2} (c), we observe that increasing the dissipative coupling to $\Omega_\mathrm{Im}/\Omega_\mathrm{Re}=1.0$ slightly shifts in energy the dispersion of the polariton branches but preserves the standard quadratic profile, except for a slight deformation of the MB and LB dispersion. This is more dramatic for the MB. While a curvature change appears at higher momenta in the MB, leading to a small negative effective mass, the LB exhibits a curvature change near zero momentum. This is clearly seen in the inset, where the change in curvature is more pronounced. There, the system's ground state becomes unstable. Accordingly, in Fig.~\ref{fig:2} (d), we observe the polaron-polariton branches as a function of detuning, where the effect of the dissipative coupling produces level attraction between the LB and MB as one moves from negative to positive detuning.

Then, we increase further the dissipative coupling to $\Omega_{\text{Im}}/\Omega_{\text{Re}}=1.5$. As seen in Fig.~\ref{fig:2} (e), the deformation over the dispersion relations of the LB and MB becomes more pronounced. Now, the curvature change in the dispersion relation of the MB is more noticeable at large momenta, as has been observed experimentally, while the LB retains quadratic behavior but inverted. As a function of detuning, this is accompanied by level attraction with different degrees between the MB and the LB or UB branches as a function of detuning, as shown in  Fig.~\ref{fig:2} (f). Ultimately, this level attraction will become an EP.

%%%%%%%%%%%%%%%%%%%%%%%%%%%%%%%%%%%%  
\subsection{Exceptional points}
%%%%%%%%%%%%%%%%%%%%%%%%%%%%%%%%%%%%

Here, we establish conditions for the emergence of EPs in the presence of the medium.  There are two sets of EPs, depending on whether the LB and MB coalesce or the MB and the UB do: $E_{\mathbf{k}}^{\text{MB}}=E_{\mathbf{k}}^{\text{LB},\text{UB}}$. This leads to:
\begin{gather}
\delta_\mathbf{k}=\Re[\Sigma_{\uparrow}(\mathbf{k},E_{\mathbf{k}}^{\text{MB}})]\pm 2\Omega_\mathrm{Im}, \\ \nonumber
\Delta\gamma=\Im[\Sigma_{\uparrow}(\mathbf{k},E_{\mathbf{k}}^{\text{MB}})]\mp 2\Omega_\mathrm{Re},
\end{gather}
or, explicitly
\begin{gather}
\delta_\mathbf{k}=\frac{\frac{4\pi n_{\downarrow}}{m_{x}}\ln[\frac{|\epsilon_{B}|}{-E_{\mathbf{k}}^{\text{MB}}+\frac{\mathbf{k}^{2}}{4m_{x}}}]}{\left[\ln\left(\frac{|\epsilon_{B}|}{-E_{\mathbf{k}}^{\text{MB}}+\frac{\mathbf{k}^{2}}{4m_{x}}}
\right)\right]^{2}+\pi^{2}\Theta(E_{\mathbf{k}}^{\text{MB}}-\frac{\mathbf{k}^{2}}{4m_{x}})}\pm 2\Omega_\mathrm{Im},\\ \nonumber
\Delta\gamma=\frac{\frac{4\pi n_{\downarrow}}{m_{x}}\pi\Theta(E_{\mathbf{k}}^{\text{MB}}-\frac{\mathbf{k}^{2}}{4m_{x}})}{\left[\ln\left(\frac{|\epsilon_{B}|}{-E_{\mathbf{k}}^{\text{MB}}+\frac{\mathbf{k}^{2}}{4m_{x}}}
\right)\right]^{2}+\pi^{2}\Theta(E_{\mathbf{k}}^{\text{MB}}-\frac{\mathbf{k}^{2}}{4m_{x}})}\mp 2\Omega_\mathrm{Re},\\
\end{gather}
We notice that, for finite momentum $\mathbf{k}$, we always have a couple of symmetric EPs for a given value of the relevant parameters ($\delta,\Omega_{\text{Im}},\Delta\gamma$).

In Fig.~\ref{fig:3}, we study the onset of EPs as a function of the relative decay $\Delta\gamma$, where we fix $\Omega_{\text{Im}}/\Omega_{\text{Re}}=1.6$. We include the standard polaron-polariton branches as dashed curves to make it easier to observe level attraction and deformation over the branches. Shifting the relative decay from negative values to zero makes to emerge EPs between the LB and MB, as shown in Figs.~\ref{fig:3} (a) to (c). However, as one approaches the positive values of $\Delta\gamma$, i.e., when the photon decay is larger than the exciton one, the EPs appear now between the UB and MB [see Fig.~\ref{fig:3} (d)]. Increasing the relative decay makes the LB and MB attract each other, and ultimately, a second set of EPs develops between them, coexisting with those from the UB and MB, as seen in Figs.~\ref{fig:3} (e) and (f).

Increasing the medium's density leads to richer behavior. In Fig.~\ref{fig:4}, we have selected a higher BEC density value of $n_{\downarrow}=9.2\times10^{10} \mathrm{cm}^{-2}$ ($4\pi n_{\downarrow}/m_{x}=1.0$) and a specific set of parameters with large exciton decay. In Fig.~\ref{fig:4} (a), we observe simultaneous level attraction between all branches for a relatively small value of dissipative coupling. This is accompanied by the corresponding deformation of the dispersion relations, as depicted in Figs.~\ref{fig:4} (b)-(d) for different values of the zero-momentum light-matter detuning, $\delta=0,\pm 2\Omega_{\text{Re}}$. This exhibits the potential of polaron-polaritons as a playground for exploiting topological changes in the spectrum of many-body excitations. 

%%%%%%%%%%%%%%%%%%%%%%%%%%%%%%%%%%%%%%%%%%%%%%%%%%%%%%%%%%%%
%%%%%%%%%%%%%%%%%%%%%%%%%%%%%%%%%%%%%%%%%%%%%%%%%%%%%%%%%%%%
\section{Discussion and Conclusions}
\label{sec:5}
%%%%%%%%%%%%%%%%%%%%%%%%%%%%%%%%%%%%%%%%%%%%%%%%%%%%%%%%%%%%
%%%%%%%%%%%%%%%%%%%%%%%%%%%%%%%%%%%%%%%%%%%%%%%%%%%%%%%%%%%%

We have studied the formation of polaron-polariton quasiparticles from an exciton-polariton impurity interacting with a polaritonic BEC in the presence of light-matter dissipative coupling. Non-Hermitian features, such as level attraction, anomalous dispersion relations (negative mass), and exceptional points, arise as a combination of the light-matter detuning, relative decay, and dissipative coupling. We have found that the lower polaron-polariton branch, associated with the coupling of light with the attractive polaron, exhibits changes of curvature for zero momentum. In contrast, the MB exhibits more dramatic changes in its dispersion relation at higher momentum. Instead, the upper polaron-polariton branch, associated with the repulsive polaron, remains mostly unaltered except for the presence of exceptional points.

When the middle polaron-polariton branch coalesces with either the lower or upper branches, one finds exceptional points. They are ultimately responsible for the deformations in the dispersion relations. A striking result is that more than one set of EPs can coexist depending on which branches coalesce, and that this can be tuned via the relative decay between the excitons and photons.

Increasing the medium's density enables on-demand level attraction, positioning polariton polarons as a valuable tool for combining non-Hermitian and many-body physics. Moreover, the emergence of novel polariton platforms to implement dissipative coupling, such as crystal waveguides~\cite{Gianfrate2024}, periodic potential~\cite{Baboux2018}, perovskite-based microcavities~\cite{Kedziora2024,Xing2026arXiv}, bilayers~\cite{Genco2025}, or hybrid organic-inorganic systems~\cite{Dutta2025}, promises a fertile playground to create cutting-edge many-body applications exploiting EPs and other non-Hermitian effects.

%%%%%%%%%%%%%%%%%%%%%%%%%%%%%%%%%%%%%%%%%%%%%%%%%%%%%%%%%%%%
%%%%%%%%%%%%%%%%%%%%%%%%%%%%%%%%%%%%%%%%%%%%%%%%%%%%%%%%%%%%
\begin{acknowledgments} 
The authors acknowledge financial support from CONAHCYT/Secihti No. CBF2023-2024-1765. A. J. V. C. acknowledges support from the Graduate Program scholarship from CONAHCYT/Secihti. M. A. B. M. acknowledges financial support from the PIPAIR 2024 project from the DAI UAM, and the Marcos Moshinsky Fellowship. A.C.G acknowledges support from PIIF25 and UNAM DGAPA PAPIIT Grant No. IA101325.
\end{acknowledgments}
%%%%%%%%%%%%%%%%%%%%%%%%%%%%%%%%%%%%%%%%%%%%%%%%%%%%%%%%%%%%
%%%%%%%%%%%%%%%%%%%%%%%%%%%%%%%%%%%%%%%%%%%%%%%%%%%%%%%%%%%%

%%%%%%%%%%%%%%%%%%%%%%%%%%%%%%%%%%%%%%%%%%%%%%%%%%%%%%%%%%%%
\appendix
%%%%%%%%%%%%%%%%%%%%%%%%%%%%%%%%%%%%%%%%%%%%%%%%%%%%%%%%%%%%

%%%%%%%%%%%%%%%%%%%%%%%%%%%%%%%%%%%%%%%%%%%%%%%%%%%%%%%%%%%%
%%%%%%%%%%%%%%%%%%%%%%%%%%%%%%%%%%%%%%%%%%%%%%%%%%%%%%%%%%%%
\section{Many-body theory}
\label{app:1}

We calculate the self-energy of the impurity exciton $\uparrow$ propagating in the BEC of $\downarrow$ excitons using the standard ladder approximation. It is expressed as
\begin{gather}
\Sigma_{\uparrow}(k)=n_{\downarrow}\mathcal{T}(k).
\end{gather}
where $k=(\mathbf{k},i\omega_{u})$ is the quadrimomentum, $\mathbf{k}$ is the momentum, and $i\omega_{u}$ is a bosonic Matsubara frequency. Here, $n_{\downarrow}$ is the density of the $\uparrow$ polariton BEC and
 \begin{gather}\label{eq:a2}
    \mathcal{T}(k)=g\left[1-g\,\Pi_{\uparrow\downarrow}(k)\right]^{-1}
\end{gather}
is the many-body scattering matrix or T-matrix. This accounts for the process in which a $\uparrow$ exciton repeatedly interacts with, and scatters a $\downarrow$ exciton out of the BEC. 
\begin{gather}\label{eq:a1}
    \Pi_{\uparrow\downarrow}(k)=-\frac{1}{\beta\mathcal{V}}\sum_{q}G_\uparrow(k+q)G_\downarrow(-q),
\end{gather}
is the propagator of the pair, $q=(\mathbf{q},i\omega_{v})$ is a quadrimomentum, $i\omega_v$ is bosonic Matsubara frequencies, $\beta=(k_BT)^{-1}$, with $k_B$ is the Boltzmann constant, $T$ is the temperature, and $\mathcal{V}$ is the 2D-volume of the semiconductor. Here, we reduce the exciton propagation in Eq.~\ref{eq:a1} to the free (non-interacting) propagators, $G_{0\sigma}(k)=i\omega_{u}-\xi_{\sigma,\mathbf{k}}^{x}$, and $\xi_{\sigma,\mathbf{k}}^{x}=\epsilon_{\mathbf{k}}^{x}-\mu_{\sigma}$ where $\mu_{\sigma}$ is the chemical potential of the $\sigma$ exciton. With this, we obtain
\begin{gather}
    \Pi_{\uparrow\downarrow}(k)=-\frac{1}{\beta V}\sum_{\mathbf{q},i\omega_v}\frac{(-i\omega_v-\xi^{x}_{\downarrow,-\mathbf{q}})^{-1}}{i\omega_u+i\omega_v-\xi^{x}_{\uparrow,\mathbf{k}+\mathbf{q}}},
\end{gather}
After replacing the momentum sum by an integral and carrying out the Matsubara sum, the expression for the pair propagator becomes
\begin{gather}
\Pi_{\uparrow\downarrow}(k)=\int\frac{\mathrm{d}^{2}\mathbf{q}}{(2\pi)^2}\frac{1+n_\mathrm{B}(\xi^{x}_{\uparrow,\mathbf{k}+\mathbf{q}})+n_\mathrm{B}(\xi^{x}_{\downarrow,-\mathbf{q}})}{i\omega_u-\xi^\uparrow_{\mathbf{k}+\mathbf{q}}-\xi^\downarrow_{-\mathbf{q}}}.
\end{gather}
where $n_\mathrm{B}(x)=\left(e^{\beta x}-1\right)^{-1}$ is the Bose-Einstein distribution function. In the zero temperature limit, by considering the chemical potential of the impurity $\mu_{\uparrow}\rightarrow\infty$, and performing the change of variables $\mathbf{q}=\mathbf{p}-\mathbf{k}/2$, one gets 
\begin{gather}
\Pi_{\uparrow\downarrow}(k)=\int\frac{\mathrm{d}^{2}\mathbf{p}}{(2\pi)^2}\frac{1}{i\omega_u-\mathbf{k}^{2}/4m_{x}-\mathbf{p}^{2}/m_{x}}.
\end{gather}
Hence, the pair propagator can be reduced to the following integral expression
\begin{gather}
\Pi_{\uparrow\downarrow}(\mathbf{k},\omega)=\frac{m_{x}}{(2\pi)^{2}}\int_{0}^{\Lambda}\int_{0}^{2\pi}\frac{du\,d\theta}{\tilde{\omega}-u+i0^{+}}.
\end{gather}
where $u=|\mathbf{p}|^{2}/2m_{x}$, $v=|\mathbf{k}|^{2}/4m_{x}$, and $\tilde{\omega}=\omega-v$. Using the identity $(x\pm i0^{+})^{-1}=\mathcal{P}(x^{-1})\mp i \pi\delta(x)$, where $\mathcal{P}$ is the principal part, we obtain
\begin{gather}
\Pi_{\uparrow\downarrow}(\mathbf{k},\omega)=\frac{m_{x}}{4\pi}\int_{0}^{\Lambda}du\left[]
\mathcal{P}\left(\frac{du}{\tilde{\omega}-u}\right)-i\pi\delta(\tilde{\omega}-u)
\right].
\end{gather}
This can be integrated analytically:
\begin{gather}
\Pi_{\uparrow\downarrow}(\mathbf{k},\omega)=-\frac{m_{x}}{4\pi}\left[\ln{\left(
\frac{\Lambda-\omega+\frac{\mathbf{k}^{2}}{4m_{x}}
}{-\omega+\frac{\mathbf{k}^{2}}{4m_{x}}}
\right)}+\right.\\ \nonumber
\left.i\pi \Theta\left(\omega-\frac{\mathbf{k}^{2}}{4m_{x}}\right)\Theta\left(\Lambda-\omega+\frac{\mathbf{k}^{2}}{4m_{x}}\right)\right]
\end{gather}
The two-body interaction in Eq.~\ref{eq:a2} is expressed in terms of the energy of the bound-state $\epsilon_\mathrm{B}<0$, called biexciton, as $g=\Re\Pi_{\uparrow\downarrow}(\mathbf{0},\epsilon_\mathrm{B})$. Because
\begin{gather}
\Pi_{\uparrow\downarrow}(0,\epsilon_{B})=-\frac{m_{x}}{4\pi}\left[\ln{\left(
\frac{\Lambda-\epsilon_{B}
}{-\epsilon_{B}}
\right)}+i\pi\Theta(\epsilon_{B})\Theta(\Lambda-\epsilon_{B})\right],
\end{gather}
It reads as follows
\begin{gather}
g=\frac{m_{x}}{4\pi}\left[\ln{\left(
\frac{-\epsilon_{B}
}{\Lambda-\epsilon_{B}}
\right)}\right].
\end{gather}
Hence, we have 
\begin{gather}
\Re\Pi_{\uparrow\downarrow}(0,\epsilon_{B})-\Re\Pi_{\uparrow\downarrow}(\mathbf{k},\omega)=\frac{m_{x}}{4\pi}\ln\left(\frac{-\epsilon_{B}}{-\omega+\frac{\mathbf{k}^{2}}{4m_{x}}}\right)
\end{gather}
and 
\begin{gather}
\Im\Pi_{\uparrow\downarrow}(\mathbf{k},\omega)=-\frac{m_{x}}{4\pi}\pi\Theta\left(\omega-\frac{\mathbf{k}^{2}}{4m_{x}}\right),
\end{gather}
where we have used $\Lambda\gg 1$. Separating the real and imaginary parts in the self-energy $\Sigma_{\uparrow}(\mathbf{k},\omega)$ we get
By substituting the pair propagator expressions and canceling the terms proportional to $\Lambda$ out, the self-energy finally becomes
\begin{gather}
\Re\Sigma(\mathbf{k},\omega)=\frac{(4\pi n_{\downarrow} /m_{x})[\ln(-\epsilon_{B})-\ln(-\omega+\frac{\mathbf{k}^{2}}{4m_{x}})]}{[\ln(-\epsilon_{B})-\ln(-\omega+\frac{\mathbf{k}^{2}}{4m_{x}})]^{2}+\pi^{2}\Theta(\omega-\frac{\mathbf{k}^{2}}{4m_{x}})},\\
\Im\Sigma(\mathbf{k},\omega)=\frac{
(4\pi n_{\downarrow} /m_{x})\pi\Theta(\omega-\frac{\mathbf{k}^{2}}{4m_{x}})
}{[\ln(-\epsilon_{B})-\ln(-\omega+\frac{\mathbf{k}^{2}}{4m_{x}})]^{2}+\pi^{2}\Theta(\omega-\frac{\mathbf{k}^{2}}{4m_{x}})}
\end{gather}

%%%%%%%%%%%%%%%%%%%%%%%%%%%%%%%%%%%%%%%%%%%%%%%%%%%%%%%%%%%%
%%%%%%%%%%%%%%%%%%%%%%%%%%%%%%%%%%%%%%%%%%%%%%%%%%%%%%%%%%%%

%%%%%%%%%%%%%%%%%%%%%%%%%%%%%%%%%%%%%%%%%%%%%%%%%%%%%%%%%%%%
%%%%%%%%%%%%%%%%%%%%%%%%%%%%%%%%%%%%%%%%%%%%%%%%%%%%%%%%%%%%
\bibliography{references}
%%%%%%%%%%%%%%%%%%%%%%%%%%%%%%%%%%%%%%%%%%%%%%%%%%%%%%%%%%%%
%%%%%%%%%%%%%%%%%%%%%%%%%%%%%%%%%%%%%%%%%%%%%%%%%%%%%%%%%%%%

\end{document}